\documentclass[aps,pra,twocolumn,superscriptaddress,notitlepage,nofootinbib,longbibliography,colorlinks=true, allcolors=blue]{revtex4-2}
\usepackage{mathrsfs}
\usepackage{epsfig}
\usepackage{graphicx}
\usepackage{amsfonts}
\usepackage{amssymb}
\usepackage{amsmath}
\usepackage{dcolumn}
\usepackage{bm}
\usepackage{xcolor}
\usepackage{braket}
\usepackage{units}
\usepackage{xspace}

\DeclareMathOperator{\sech}{sech}
\usepackage{bbold}
\usepackage{orcidlink}
\usepackage{tabularx}
\usepackage{adjustbox}
\usepackage{booktabs}
\usepackage[]{hyperref}
\newcommand*\midpoint[1]{\overline{#1}}
\newcommand{\JSU}{Department of Physics, Jiangsu University, Zhenjiang 212013, China}
\newcommand{\ECNU}{Quantum Institute for Light and Atoms, State Key Laboratory of Precision Spectroscopy, Department of Physics, School of Physics and Electronic Science, East China Normal University, Shanghai 200062, China}
\newcommand{\KU}{Mathematics Department, Khalifa University of Science and Technology, Abu Dhabi 127788, United Arab Emirates}
\newcommand{\ZJNU}{Department of Physics, Zhejiang Normal University, Jinhua 321004, China}

\begin{document}
\title{%
Compasslike states in a thermal reservoir and fragility of their nonclassical features}

\author{Naeem Akhtar}
\affiliation{\JSU}

\author{Xiaosen Yang}
\email{yangxs@ujs.edu.cn}
\affiliation{\JSU}

\author{Muhammad Asjad}
\affiliation{\KU}

\author{Jia-Xin Peng}
\affiliation{\ECNU}

\author{Gao Xianlong}
\email{gaoxl@zjnu.edu.cn}
\affiliation{\ZJNU}

\author{Yuanping Chen}
\email{chenyp@ujs.edu.cn}
\affiliation{\JSU}
\date{\today}
\begin{abstract}

Superposed photon-added and photon-subtracted squeezed-vacuum states exhibit sub-Planck phase-space structures and metrological potential similar to the original compass states (superposition of four coherent states), but are more closely tied to modern experiments. Here, we observe that these compasslike states are highly susceptible to loss of quantum coherence when placed in contact with a thermal reservoir; that is, the interaction with the thermal reservoir causes decoherence, which progressively suppresses the capacity of these states to exhibit interference traits. We focus on the sub-Planck structures of these states and find that decoherence effects on these features are stronger with increasing the average thermal photon number of the reservoir, the squeezing parameter, or the quantity of added (or subtracted) photons to the squeezed-vacuum states. Furthermore, we observe that the sub-Planck structures of the photon-subtracted case survive comparatively longer in the thermal reservoir than their counterparts in the photon-added case, and prolonged contact with the thermal reservoir converts these compasslike states into a classical state.
\end{abstract}
\maketitle

\section{Introduction}\label{sec:introduction}

The prototypical Schr\"{o}dinger cat state, a superposition of macroscopically distinct quantum states, originated from the well-known \textit{gedanken experiments}~\cite{Sch35}, in which a cat appeared to be both alive and dead simultaneously. This notion morphed into a macroscopic cat state~\cite{Mil86,YS86}, which is the superposition of two distinguishable coherent states, and then into multicomponent cat states~\cite{Zurek2001,nam2018}. Such states are of great interest because they may hold particular non-classical features such as non-Gaussian interference characteristics~\cite{Qin2021,Schleich1991} and sub-Planck phase-space structures~\cite{Zurek2001}, and have both fundamental and practical implications in the field of continuous-variable quantum information processing. These implications span from understanding the fundamental physics of quantum decoherence~\cite{Collapse2} to practical implications in quantum metrology~\cite{Joo2011,pan2014,mitchell2004}, quantum teleportation, and cryptography~\cite{van2001,Jeong2001,Brask2010}, to name a few.

The addition (or subtraction) of photons to the squeezed-vacuum states (SVSs) leads to the quantum states having Wigner phase-space features similar to those of cat states~\cite{Meng12,TANG201586,zhang1992}. Similarly, when photons are added to or subtracted from the superposition of a SVS, one may obtain compasslike states~\cite{naeem2023} that have comparable sub-Planck phase-space structures as the original compass states~\cite{Zurek2001}.\;Sub-Planck structures have
been explored in various contexts~\cite{Eff1,Eff3,Eff8,Eff9,Eff10,panigrahi2022compass,Howard2019,Atharva2023,Naeem2022,Naeem2021}, and it has been found that such structures are highly sensitive to environmental decoherence~\cite{Zurek2003} and play a crucial
role in the sensitivity of a quantum state against phase-space displacements~\cite{Facon2016,Toscano06,Eff4}.\;Both theoretically and experimentally, the addition (or subtraction) of photons to the SVSs has been explored~\cite{Dakna1997,Takase2021,Prop6,Alexei2006,Neergaard2006}, and it has been found that these methods are quite effective in producing larger cat states for quantum computing~\cite{Thekkadath2020}.

The interaction of a quantum system with its surrounding environment leads to the loss of its quantum features, a so-called decoherence phenomena~\cite{Collapse2,RevModPhys.90.025004}. Superposed states, such as macroscopic catlike states, are theoretically attainable but difficult to achieve in practical applications due to their extreme sensitivity to environmental decoherence~\cite{Agarwal1971,gardiner2004}. In the case of macroscopic systems, the interaction with the environment can never be avoided because the decoherence rate is proportional to the ‘‘macroscopic separation’’ between the two states~\cite{zurek19911,Caldeira1985,Walls1985,Jeong2007}, and when propagating through damping channels, such states rapidly lose their non-classical properties and the corresponding negative oscillations of their Wigner functions~\cite{Kim1993,dodonov2016,Hu2010,Hu2010b,kim1992,bera2023fragility,deleglise2008}.\;Consequently, it has been discovered that although catlike states are theoretically possible, they are difficult to observe in actual experiments as a macroscopic object like a cat cannot be completely isolated from its surroundings~\cite{wheeler2014}.\;The recent rapid advancement in the theory of quantum information processing, which involves the protection of quantum states and their quantum dynamics after contact with an environment generating decoherence, is a highly important subject~\cite{Jeannic2018,Pan2023,Vitali1998}.

In this work, we theoretically investigate the interaction between the quantum states of our interest and a thermal reservoir that acts as a cause of decoherence; in particular, we examine how the decoherence deforms the non-classical phase-space assets of our states. The Fokker-Planck equation that controls the temporal evolution of the relevant quantum system can be obtained directly from the master equation for the density operator in the Born-Markov approximations~\cite{Agarwal1971,gardiner2004}.\;Using the Wigner quasidistribution formalism~\cite{Sch01}, we solve the Fokker-Planck equations to obtain the temporal evolution of the associated Wigner functions under a thermal reservoir~\cite{gardiner2004}.\;We pick compasslike states for this interaction, which were recently obtained by adding to or subtracting photons from two superposed SVSs~\cite{naeem2023}.\;As previously mentioned, the Wigner functions of these states contain sub-Planck structures, which makes them a potential substitute for the original compass states~\cite{Zurek2001}.\;Moreover, compared to the original compass states, these states may be more appropriate for contemporary experiments~\cite{Alexei2006,Neergaard2006}.

We mainly focus on analyzing the impact of the given thermal reservoir on the sub-Planck structures of the stated compasslike states.\;Visualizing plots of the corresponding Wigner functions, we observe that the sub-Planck structures of these compasslike states are highly influenced by the thermal reservoir; that is, the decoherence aroused by this interaction may eventually smear out these structures from the phase space. A numerical examination of these Wigner function plots reveals that the corresponding sub-Planck structures decay more quickly as the average thermal photon number of the reservoir, the quantity of added (or subtracted) photons to the squeezed-vacuum states, or the squeezing parameter increase.\;Compared to the corresponding photon-subtracted case, the sub-Planck structures related to the photon-added case decay faster in the thermal reservoir.\;Furthermore, we observe that the long-term contact of these compasslike states with the thermal reservoir eventually converts them into a thermal state.

The structure of our paper is as follows: \S\ref{sec:background} contains an overview of the basic concepts that are employed throughout this manuscript.\;We address the interaction between the quantum states of our interest and a thermal reservoir in \S\ref{sec:inter}. Finally, \S\ref{sec:conc} contains the main conclusions and the detailed physical explanations of our findings.

\section{Theoretical framework}\label{sec:background}
This section provides an overview of the main ideas and findings from earlier research that are relevant to the present work. We organized this section as follows: First, we review the basic concept of the phase-space representation of a quantum state via the Wigner function in \S\ref{subsec:Wigner_fucntion}, and then this concept is further reviewed for the quantum states of our interest in \S\ref{subsec:non_Gaussian_states}. In \S\ref{subsec:models}, we revisit the thermal reservoir that will be employed to interact with the quantum states of \S\ref{subsec:non_Gaussian_states}.

\subsection{Phase-space analysis}\label{subsec:Wigner_fucntion}
The Wigner quasiprobability distribution is the visualization of quantum mechanical states in phase space~\cite{Sch01,Wig32,royer1977,leonhardt1997measuring}. The Wigner function for a generic quantum state
$\hat{\rho}$ is written as an expectation value of the parity kernel~\cite{royer1977},
\begin{align}
    W\left(\alpha\right)
    :=\text{tr}\left[\hat{\rho}\hat{\Delta}(\alpha)\right]~\text{with}~\alpha\in\mathbb{C},
\label{eq:wigner_general}
\end{align}
where
\begin{align}
 \hat{\Delta}(\alpha):=2\hat{D}(\alpha)\hat{\Pi}\hat{D}^\dagger(\alpha),\,
 \hat{\Pi}
:=\left(-1\right)^{\hat{a}^\dagger\hat{a}}
\end{align}
is the displaced parity operator, and $\hat{a}^{\dagger}$ ($\hat{a}$) are raising (lowering) operators that satisfy the commutation relation $[\hat{a},\hat{a}^\dagger]=1$.

The quantum uncertainty principle~\cite{leonhardt1997measuring,Robertson1929,wheeler2014}
arising from commutator relations $\left[\hat{x},\hat{p}\right]=\text{i}$ of the position operator $\hat{x}$ and the momentum operator $\hat{p}$ obeys
\begin{align}
\label{eq:uncertain}
\Delta x\Delta p\geq\frac{1}{2},
\end{align}
here
\begin{equation}
\label{eq:DeltaA2}
\Delta C^2:=\braket{\hat{C}^2}-\braket{\hat{C}}^2
\end{equation}
is the uncertainty of any operator~$\hat{C}$, and we set $\hbar= 1$ hereafter. Hence, note that in the following, we use dimensionless
versions of position and momentum operators.

The single-mode SVS can be written as
\begin{align}
\ket{\psi}=\hat{S}(r)\ket{0}
\end{align}
with
\begin{align}
\hat{S}(r):=&\exp\bigg[\frac{r}{2}\left(\hat{a}^{\dagger2}-\hat{a}^2\right)\bigg],
\end{align}
is the squeezing operator~\cite{CarlosNB15}, with $r$ being the real squeezing parameter. The Wigner function of the SVS is
\begin{align}
W(\alpha)=\frac{2}{\pi}\exp\left(-2|\Bar{\alpha}|^2\right),
\end{align}
where
\begin{align}
\Bar{\alpha}=\alpha \cosh (r)-\alpha^*\sinh (r)~\text{with}~\alpha:=\nicefrac{(x+\text{i}p)}{\sqrt{2}}.
\end{align}
The Wigner function exhibits a Gaussian distribution, indicating that the SVS is a Gaussian state that deviates from classicality when these states are squeezed~\cite{CarlosNB15}.
Non-Gaussian operations such as photon addition (PA) or subtraction (PS) applied to a Gaussian SVS lead to non-Gaussian SVSs~\cite{zhang1992,Biswas2007,Dey2020,Meng12}; that is, the Wigner functions of the resulting states may attain non-Gaussian terms. The non-classical nature of such states is indicated by the negative value of their Wigner function, apart from the squeezing in their quadrature, which is another indicator of the aforementioned nature.

The sub-Planck structures are the phase-space features having dimensions far smaller than the bound given by the uncertainty relation (\ref{eq:uncertain}), and such structures have achieved significant attention in quantum metrology~\cite{Facon2016,Toscano06,Eff4} due to their high sensitivity to environmental decoherence~\cite{Zurek2003}. The PA and PS cases of the superposed SVSs have been previously analyzed, and the Wigner functions of these states are shown to have sub-Planck structures~\cite{naeem2023}. Specifically, it has been found that an excess amount of the photons added to or subtracted from the superposition of SVSs can cause sub-Planck structures in the phase space. We revisit the sub-Planck structures contained by the Wigner functions of these non-Gaussian SVSs next.

\begin{figure*}
\centering
\includegraphics[width=1.02\textwidth]{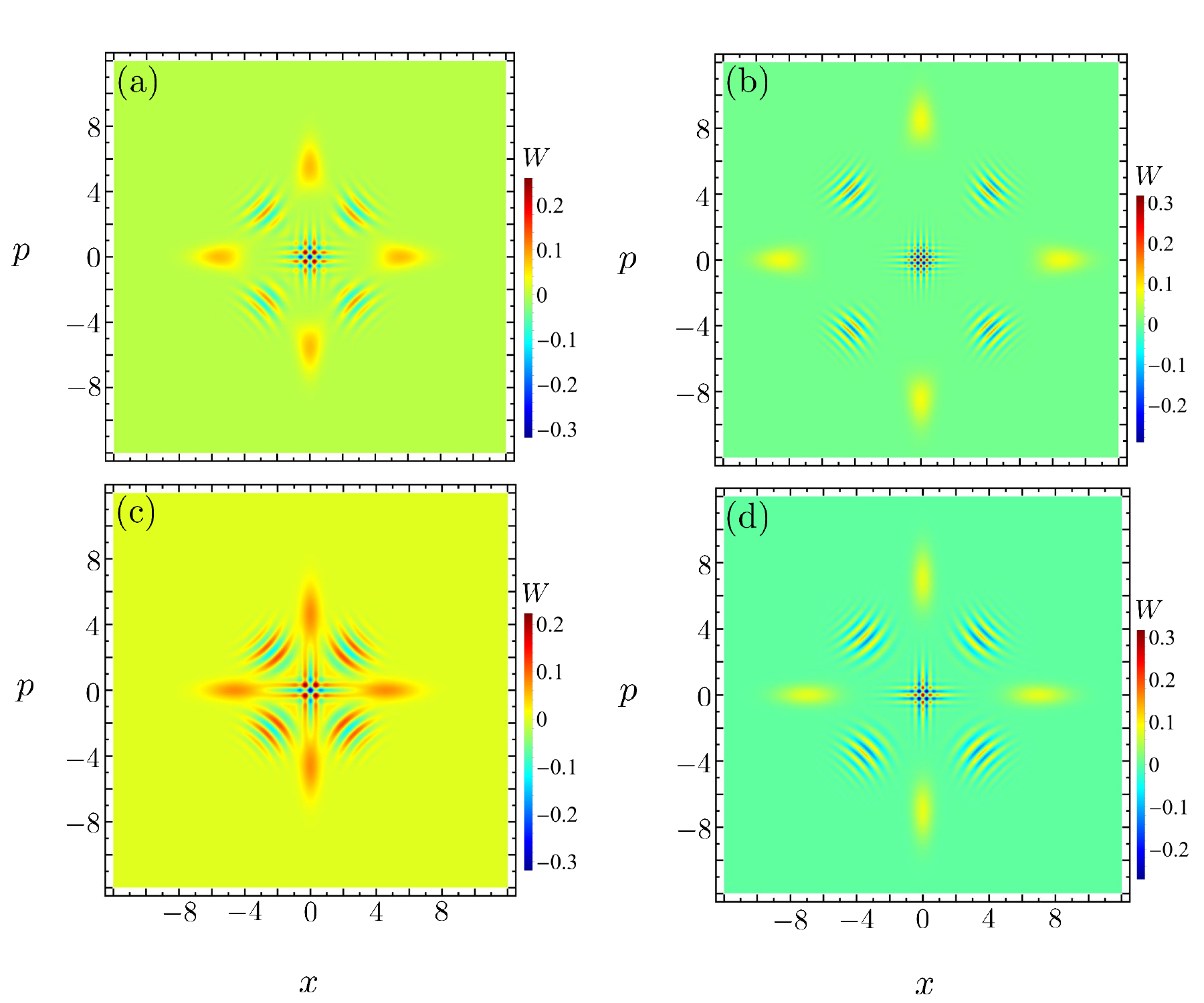}
\caption{The Wigner function of the PA and PS cases of the superposed SVSs: the corresponding PA cases with (a) $m=5$ and (b) $m=12$, and the corresponding PS cases with (c) $m=5$ and (d) $m=12$. For all cases, $r=0.8$.}
\label{fig:fig1}
\end{figure*}

\subsection{Compasslike states}\label{subsec:non_Gaussian_states}
This section provides an overview of the recently introduced compasslike states~\cite{naeem2023}, which were obtained by either adding photons to or subtracting photons from the superposition of two SVSs. It has been shown that when a significant quantity of photons is added (or subtracted), the Wigner functions of these states are shown to have sub-Planck phase-space features. Let us first review the case when $m$ photons are added to the superposed SVSs, which can be expressed as follows:

 \begin{align}\label{eq:PAsvs}
\ket{\psi_{\text{PA}}}:=N_\text{PA}\hat{a}^{\dagger m}\left(\hat{S}(r)\ket{0}+\hat{S}(-r)\ket{0}\right),
\end{align}
where
\begin{align}
N_\text{PA}:=&\nonumber\Bigg[\frac{2}{\sqrt{\cosh(2r)}}\left(\frac{\cosh(r)}{\sqrt{\cosh(2r)}}\right)^{m}m!P_{m}\left(\frac{\cosh(r)}{\sqrt{\cosh(2r)}}\right)
\\&+2\cosh^{m} (r) m!P_{m}\big(\cosh(r)\big)
\Bigg]^{-\frac{1}{2}},
\end{align}
is the normalization coefficient. 

The Wigner function of the state $\ket{\psi_{\text{PA}}}$ is calculated as~\cite{naeem2023}
\begin{align}\label{eq:wig_spa}
W_{\ket{\psi_{\text{PA}}}}(\alpha)=N^2_\text{PA}\left[W_{\bigoplus}(\alpha)+2 \operatorname{Re}\big(W_{\bigotimes}(\alpha)\big)\right].
\end{align}
The first term is
\begin{align}\label{eq:pa_org_Wig1}
W_{\bigoplus}(\alpha)=W^+(\alpha)+W^-(\alpha),
\end{align}
where
\begin{align}
    W^{\pm}(\alpha)
    =&\nonumber\aleph_{\pm}\sum_{l=0}^{m}\frac{\left(m!\right)^{2}\left[\mp 2 \coth (r)\right]^{l}}{l!\left[(m-l)!\right]^{2}}\\&\times\left|H_{m-l}\left[-\mathrm{i}\sqrt{\pm 2\coth(r)}\bar{\alpha}_{\pm}\right]\right|^2
    \label{eq:expr_PASV1}
\end{align}
with $H_m$ representing the Hermite polynomial, and
\begin{align}
\aleph_{\pm}:=\frac{\text{e}^{\theta_{\pm }}\left[\pm \sinh(2r)\right]^m}{\pi 2^{2m}},\bar{\alpha}_{\pm}:=\alpha \cosh (r)\mp \alpha^* \sinh(r),
\end{align}
where
\begin{align}
\theta_{\pm}:=\pm\sinh (2r)\left(\alpha^{*2}+\alpha^2\right)-2|\alpha|^2\cosh (2r).
\end{align}
The second term in Eq.~(\ref{eq:wig_spa}) is
\begin{align}\label{eq:cross_1}
&\nonumber W_{\bigotimes}(\alpha):=O^-_m\sum^m_{l=0}\frac{(m!)^2\left[-2\text{i}\coth(r)\right]^l}{[(m-l)!]^2}H_{m-l}\left[\text{i}\Omega\alpha_{-}\right]\\&\times H_{m-l}\left[-\Omega\alpha^*_+\right],O^{\pm}_m=\frac{\text{e}^{\theta}\left[\pm\text{i} \tanh(2 r)\right]^m}{\pi 2^{2m}\cosh(r)\sqrt{1+\tanh^2(r)}},
\end{align}
where
\begin{align}
\Omega:=\frac{\sqrt{\tanh(2r)}}{\sinh(r)},
\end{align}
and
\begin{align}
\theta:=-\tanh(2r)\left(\alpha^2-\alpha^{2 *}\right)-2|\alpha|^2\sech(2r)
\end{align}
with
\begin{align}
\alpha_{\pm}:=\alpha^*\sinh(r)\pm\alpha \cosh(r).
\end{align}
We plot the Wigner function of the state $\ket{\psi_{\text{PA}}}$ in Figs.~\ref{fig:fig1}(a) and \ref{fig:fig1}(b) by varying photon number $m$. This implies that as the number of added photons increases, the area of the resulting sub-Planck structures decreases. The four Gaussian-like peaks and the chessboard-like pattern are produced by the term $W_{\bigoplus}(\alpha)$, where the tiles in the chessboard pattern are sub-Planck structures in phase space. On the other hand, oscillations that occur far from the phase-space center are caused by $W_{\bigotimes}(\alpha)$.

In a similar way, subtracting $m$ photons from the superposition of two SVSs yields
\begin{align}\label{eq:PSsvs}
\ket{\psi_{\text{PS}}}:=N_\text{PS}\hat{a}^{ m}\left(\hat{S}(r)\ket{0}+\hat{S}(-r)\ket{0}\right),
\end{align}
where
\begin{align}
N_\text{PS}:=&\nonumber\Bigg[\frac{2}{\sqrt{\cosh(2r)}}\left(\frac{\sinh(r)}{\sqrt{\cosh(2r)}}\right)^{m}
m!
P_{m}\left(\frac{\sinh (r)}{\sqrt{\cosh(2r)}}\right)
\\&+2\big(-\mathrm{i}\sinh(r)\big)^{m}m!P_{m}\big(\mathrm{i}\sinh(r)\big)
\Bigg]^{-\frac{1}{2}},
\end{align}
is the normalization coefficient, and the corresponding Wigner function of this state $\ket{\psi_{\text{PS}}}$ is calculated as~\cite{naeem2023}
\begin{align}
W_{\ket{\psi_{\text{PS}}}}(\alpha)=N^2_\text{PS}\left[W_{\bigoplus}(\alpha)+2 \operatorname{Re}\big(W_{\bigotimes}(\alpha)\big)\right].
\label{eq:wig_sps}
\end{align}
With
\begin{align}
    W^\pm(\alpha)
    =&\nonumber\aleph_{\pm}\sum_{l=0}^{m}\frac{\left(m!\right)^{2}\left[\mp 2 \tanh (r)\right]^{l}}{l!\left[(m-l)!\right]^{2}}\\&\times\left|H_{m-l}\left[-\mathrm{i}\sqrt{\pm 2\tanh(r)}\bar{\alpha}_{\pm}\right]\right|^2,
\end{align}
the first term of Eq.~(\ref{eq:wig_sps}) is expressed as
\begin{align}\label{eq:chess_ps}
W_{\bigoplus}(\alpha)=W^+(\alpha)+W^-(\alpha).
\end{align}
The second term of Eq.~(\ref{eq:wig_sps}) is
\begin{align}\label{eq:secondterm_ps}
W_{\bigotimes}(\alpha):=&\nonumber O^+_m\sum^m_{l=0}\frac{(m!)^2\left[2\text{i}\tanh(r)\right]^l}{[(m-l)!]^2}H_{m-l}\left[-\omega\alpha_{-}\right]\\& \times H_{m-l}\left[-\text{i}\omega\alpha^*_+\right],\omega:=\frac{\sqrt{\tanh(2r)}}{\cosh(r)}.
\end{align}
The relevant Wigner function for the state $\ket{\psi_{\text{PS}}}$ is shown in Figs.~\ref{fig:fig1}(c) and \ref{fig:fig1}(d) for different values of $m$, and similar to the PA case, we observe that the area of the sub-Planck structures decreases when the number of subtracted photons is increased. Similarly as in the PA case, the term $W_{\bigoplus}(\alpha)$ produces Gaussian-like peaks and a central pattern resembling a chessboard consisting of sub-Planck structures in phase space. And the interference pattern that appears far from the phase-space origin is caused by the term $W_{\bigotimes}(\alpha)$. Note that an excess amount of the photons, that is, $m\gg1$, is required to induce the sub-Planck structures in the Wigner phase space of the corresponding states.

The superposed photon-added and photon-subtracted SVSs presented above can be an alternative to the original compass states~\cite{Zurek2001}, which have the same phase-space features but may have a connection with more feasible experiments. For example, it has been previously demonstrated that the superpositions of coherent states could be constructed deterministically by using third-order Kerr nonlinearity~\cite{YS86}. However, this method requires the availability of Kerr nonlinearity for an order, which is actually not applicable to currently available Kerr media. Also, states of this type are particularly prone to loss, and because absorption cannot be ignored in currently available Kerr media, the capacity to extract coherent-state superpositions before they lose their quantum properties is severely limited~\cite{Boyd1999}. Hence, adding or subtracting photons from Gaussian SVSs seems to be a more feasible approach for creating coherent-state superpositions~\cite{Alexei2006, Neergaard2006}.

 Adding (subtracting) the same amounts of photons to (from) SVSs may result in quantum states with different phase-space characteristics~\cite{Ma2012}, as illustrated in Fig.~\ref{fig:fig1} for our compasslike states, where the PA case yields smaller sub-Planck structures than the PS case~\cite{naeem2023}. Furthermore, these PA and PS cases also differ significantly in other characteristics, such as that the PA case always holds a higher average photon number~\cite{WANG2019102,Richard2014} and metrological potential~\cite{naeem2023,WANG2019102} than the PS case. The interaction of these compasslike states with a noisy environment may alter their phase-space attributes, and a prolonged interaction may destroy all of their critical quantum traits. We believe that our compasslike states may also perform differently in a noisy environment, and establishing which of these two quantum states
is better at maintaining their quantum properties against environmental degradation is an
important research topic to address.\;The topic of interaction between a quantum system and its environment has been intensively investigated since the beginning~\cite{Brune1996,Monroe1996b,Turchette1998,Poyatos1996,Leggett1987a,Leggett1987b,Caldeira1983,CALDEIRA1983587,Caldeira1985,Leggettsch1983,gardiner2004,Agarwal1971,zurek19911} and is a crucial issue to discuss~\cite{Pan2023}. In this study, we will mainly
focus on the interaction of our quantum states with a heat
reservoir, observing how the decoherence generated by this interaction modifies our states. Next, we present an overview of the heat reservoir under consideration in this study.

\subsection{Thermal channels}\label{subsec:models}
The influence of damping on the quantum properties of systems was originally discussed in~\cite{Brune1996,Monroe1996b,Turchette1998,Poyatos1996,Leggett1987a,Leggett1987b,Caldeira1983,CALDEIRA1983587,Caldeira1985,Leggettsch1983,gardiner2004,Agarwal1971,zurek19911}, and it has been found that such damping channels have a strong impact on the quantum properties held by a system~\cite{Leggettsch1983,zurek19911}.
This section discusses the model of a finite-temperature thermal reservoir, which in the present work is referred to as a thermal channel. The master equation describes the time evolution of a single-mode state denoted by the density operator $\hat{\rho}$ in this thermal channel~\cite{gardiner2004,Agarwal1971},
\begin{align}\label{eq:models_decoh}
    \frac{\text{d} \hat{\rho}}{\text{d} t}=&\nonumber\kappa \left(\Bar{n}+1\right)\left(2\hat{a}\hat{\rho}\hat{a}^{\dagger}-\hat{a}^{\dagger}\hat{a}\hat{\rho}-\hat{\rho}\hat{a}^{\dagger}\hat{a}\right)\\&+\kappa \Bar{n}\left(2\hat{a}^{\dagger}\hat{\rho}\hat{a}-\hat{a}\hat{a}^{\dagger}\hat{\rho}-\hat{\rho}\hat{a}\hat{a}^{\dagger}\right),
\end{align}
where $\kappa$ is the decay rate, and~$\Bar{n}$ denotes the average number of thermal photons in the cavity. The first term on the right-hand side of Eq.~(\ref{eq:models_decoh}) describes the transfer through the decay of photons from the quantum system to the thermal reservoir, while the second term corresponds to the transfer of excitation from the non-zero temperature thermal reservoir to the quantum system. For $\Bar{n}=0$, Eq.~(\ref{eq:models_decoh}) leads to the case describing the decay of the quantum state to a zero-temperature reservoir (also known as the photon-loss channel)~\cite{kim1992}.

The temporal evolution of the Wigner function of a quantum state in the thermal channel given by Eq.~(\ref{eq:models_decoh}) can be obtained as~\cite{Hu2010b}
\begin{align}\label{eq:tempo_wig}
 W(\zeta,t)=\frac{2}{\midpoint{T}}\int\frac{d^2\alpha}{\pi} W(\alpha)\exp\left(\frac{-2\left|\zeta-\alpha\text{e}^{-\kappa t}\right|^2}{\midpoint{T}}\right),
\end{align}
where $\zeta=\nicefrac{(x+\text{i}p)}{\sqrt{2}}$ and $\midpoint{T}=(1+2\Bar{n})T$ with $T=1-\text{e}^{-2\kappa t}$. Note that in the following, we utilized a dimensionless version of the time, $\tau=\kappa t$, for our convince.

The subsequent sections provide a theoretical analysis of the interactions between the thermal reservoir covered in this section and the compasslike states of \S\ref{subsec:non_Gaussian_states}. To determine the time evolution of the system prepared in these compasslike states, we use the exact solution of the Fokker-Planck equation described through the master equation in (\ref{eq:models_decoh}) for the Wigner quasidistributions, which are easily obtained by solving Eq.~(\ref{eq:tempo_wig}).\,This solution clearly describes the impact of the thermal channel on the quantum dynamics of the compasslike states under consideration. Here, we primarily study the decoherence effects resulting from this interaction on the sub-Planck structures of these compasslike states.

The thermal reservoir given by Eq.\,(\ref{eq:models_decoh}) has been shown to have a substantial effect on the non-classical features retained by a quantum state in phase space; that is, cavity damping causes decoherence, which swiftly eliminates the oscillations of the Wigner functions.For example, this phenomenon has been previously discussed for other quantum states as well~\cite{Kim1993,dodonov2016,Hu2010,Hu2010b,kim1992,bera2023fragility,deleglise2008}.
We now discuss the interaction of our quantum states with a thermal reservoir and explain how decoherence affects their nonclassical phase-space assets and stability in the presence of decoherence.

\section{Interaction with environment}\label{sec:inter}
In this section, we address the interaction between the compasslike states of \S\ref{subsec:non_Gaussian_states} and the thermal reservoir provided in \S\ref{subsec:models}. We discuss in detail how the interaction with the given thermal reservoir deforms the nonclassical phase-space regions contained by these compasslike states, specifically their sub-Planck structures. To accomplish this, we compute the temporal evolution of the Wigner functions associated with these states in the thermal reservoir by using Eq.~(\ref{eq:tempo_wig}), and our analysis is structured in the following sections: First, in \S\ref{subsec:pa_case_math}, we provide the mathematical terms involved in the temporal evolution of the corresponding Wigner functions of our compasslike states, and then the discussion about these temporal evolution of the Wigner functions is provided in \S\ref{subsec:tempo_pa_vs_ps}. The impact of the thermal reservoir on the sub-Planck structures contained by these compasslike states is covered in \S\ref{subsec:decay_rate}, and finally, in \S\ref{subsec:compare_pa_vs_ps}, we provide a numerical check of the results presented in \S\ref{subsec:decay_rate}.

\subsection{Temporal Evolution of Wigner functions}\label{subsec:pa_case_math}
This section presents detailed mathematics related to the temporal evolution of the corresponding Wigner function of our stated compasslike states, which will be used later in our analysis to assess the effects of the heat reservoir on these states. To make our approach more understandable, we split the temporal evolution of these Wigner functions into a number of mathematical terms.

First, consider the Wigner function of the PA case provided in Eq.~(\ref{eq:wig_spa}). The component of temporal evolution associated to its first term contains $W^\pm(\alpha)$ that is obtained by using Eq.~(\ref{eq:tempo_wig}), and it is given by
\begin{align}\label{eq:tempo_pa_chess}
W^\pm(\zeta,\tau)=&\nonumber \frac{\left[\pm \sinh(2r)\right]^m\text{e}^{\frac{B^{\pm}_1}{A^+}-\frac{2|\zeta|^2}{\midpoint{T}}}}{\pi~ 2^{2m-1}\midpoint{T}\sqrt{A^+}}\sum^m_{l=0}\sum^{m-l}_{k=0}\\&\times \nonumber\frac{(m!)^2[\mp 2\coth(r)]^l (-G^{\pm}_1)^k \chi_1^{m-l-k} }{k! l! \big[(m-l-k)!\big]^2 (-A^+)^{m-l}}\\& \times H_{m-l-k}\left(\text{i}\Theta_1C^{\pm}_1\right)H_{m-l-k}\left(\text{i}\Theta_1D^{\pm}_1\right)
\end{align}
with $A^+$, $B{_1}^\pm$, $C^{\pm}_1$, $D^{\pm}_1$, $G^{\pm}_1$, $\chi_1$, and $\Theta_1$ are provided in the Appendix \ref{appendix_b}.

Similarly, the component of the temporal evolution of the Wigner function for the second term of Eq.~(\ref{eq:wig_spa}), denoted by $W_{\bigotimes}(\alpha)$, is calculated as
\begin{align}\label{eq:tempo_pa_cross}
W_{\bigotimes}(\zeta,\tau)=&\nonumber\frac{[-\tanh(2r)]^m\text{e}^{\frac{B^+_2}{A^-}-\frac{2|\zeta|^2}{\midpoint{T}}}}{\pi \midpoint{T} \sqrt{A^-}\cosh( r)2^{2m-1}\sqrt{1+\tanh^2(r)}}\sum^m_{l=0}\\&\times \sum^{m-l}_{k=0}\nonumber\frac{(m!)^2[2\text{i}\coth (r)]^l(-G_2)^k \chi_2^{m-l-k}}{k!~l![(m-l-k)!]^2(-A^-)^{m-l}}\\&\times H_{m-l-k}\left(\text{i}\Theta_2C_2\right)H_{m-l-k}\left(\text{i}\Theta_2D_2\right),
\end{align}
where $A^-$, $B^+_2$, $C_2$, $D_2$, $G_2$, and $\Theta_2$ are given in the Appendix \ref{appendix_b}.

Similar to the PA case, the Wigner function (\ref{eq:wig_sps}) of the PS case is modified after interacting with the heat reservoir. Let us consider its first and second terms, and then describe their related temporal evolution as illustrated below. The first term of Eq.~(\ref{eq:wig_sps}) modifies as
\begin{align}\label{eq:tempo_ps_chess}
W^\pm(\zeta,\tau)=&\nonumber \frac{[\pm \sinh(2r)]^m\text{e}^{\frac{B^{\pm}_1}{A^+}-\frac{2|\zeta|^2}{\midpoint{T}}}}{\pi~ 2^{2m-1}\midpoint{T}\sqrt{A^+}}\sum^m_{l=0}\sum^{m-l}_{k=0}\\&\times\nonumber\frac{(m!)^2[\mp 2\tanh (r)]^l (-G^{\pm}_3)^k \chi_3^{m-l-k} }{k! l! \big[(m-l-k)!\big]^2 (-A^+)^{m-l}}\\& \times H_{m-l-k}\left(\text{i}\Theta_3C^{\pm}_3\right)H_{m-l-k}\left(\text{i}\Theta_3D^{\pm}_3\right)
\end{align}
with $C^{\pm}_3$, $D^{\pm}_3$, $G^{\pm}_3$, $\chi_3$, and $\Theta_3$ provided in the Appendix \ref{appendix_b}.

Similarly, for the second term of Eq.~(\ref{eq:wig_sps}), we have
\begin{align}\label{eq:tempo_ps_cross}
W_{\bigotimes}(\zeta,\tau)=&\nonumber\frac{[-\text{i}\tanh(2r)]^m\text{e}^{\frac{B^-_2}{A^-}-\frac{2|\zeta|^2}{\midpoint{T}}}}{\pi \midpoint{T} \sqrt{A_2}\cosh(r)2^{2m-1}\sqrt{1+\tanh^2(r)}}\sum^m_{l=0}\\&\times \sum^{m-l}_{k=0}\nonumber\frac{(m!)^2[2\text{i}\tanh(r)]^l(-G_4)^k \chi_4^{m-l-k}}{k!~l![(m-l-k)!]^2(-A^-)^{m-l}}\\&\times H_{m-l-k}\left(\text{i}\Theta_4C_4\right)H_{m-l-k}\left(\text{i}\Theta_4D_4\right),
\end{align}
where $B^-_2$, $C_4$, $D_4$, $\chi_4$, and $\Theta_4$ are provided in the Appendix \ref{appendix_b}.

All of the constituents described in this section for the temporal evolution of the Wigner function of the PA and PS instances will be included in our subsequent analysis.
\begin{figure*}[t]
\centering
\includegraphics[width=1.03\textwidth]{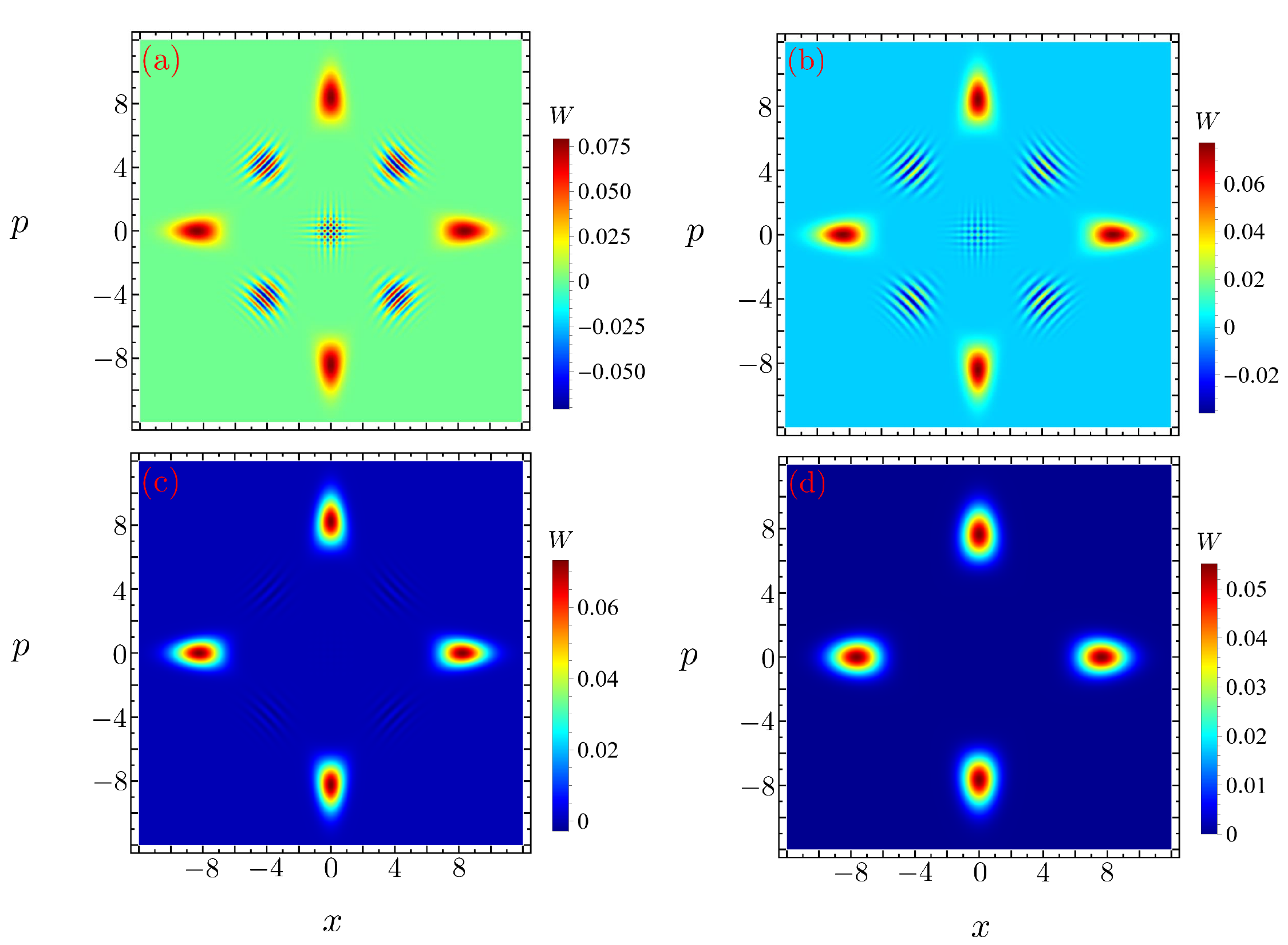}
\caption{The Wigner function of the PA case in the thermal reservoir, where $m=12$ and $r=0.8$. (a) $\Bar{n}=0$ and $\tau=0.01$; (b) $\Bar{n}=0.5$ and $\tau=0.01$; (c) $\Bar{n}=0.5$ and $\tau=0.03$; and (d) $\Bar{n}=1$ and $\tau=0.1$.}
\label{fig:fig2}
\end{figure*}
\begin{figure*}
\centering
\includegraphics[width=1.03\textwidth]{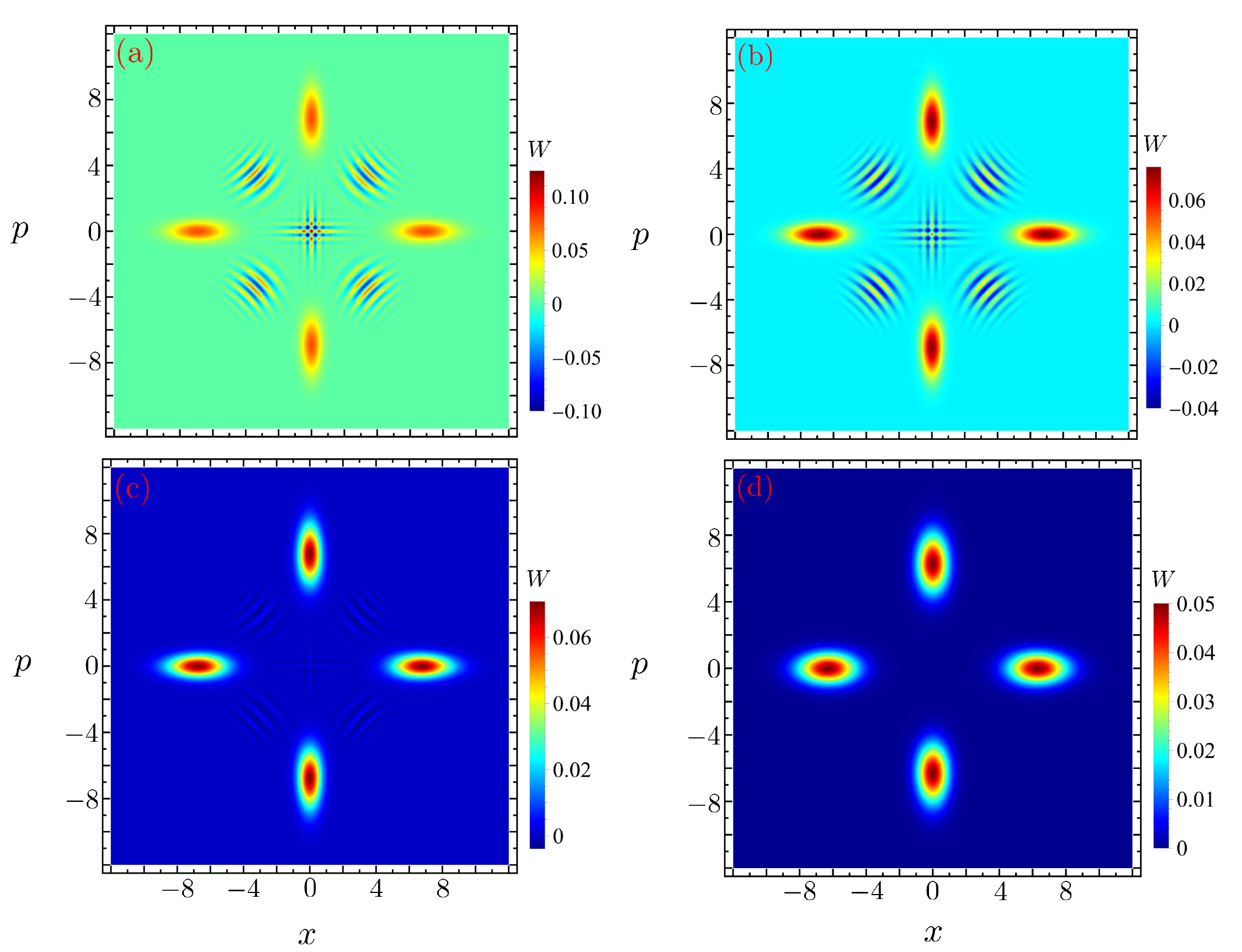}
\caption{The Wigner function of the PS case in the thermal reservoir, where $m=12$ and $r=0.8$. (a) $\Bar{n}=0$ and $\tau=0.01$; (b) $\Bar{n}=0.5$ and $\tau=0.01$; (c) $\Bar{n}=0.5$ and $\tau=0.03$; and (d) $\Bar{n}=1$ and $\tau=0.1$.}
\label{fig:fig3}
\end{figure*}
\subsection{Fragility of nonclassical features}\label{subsec:tempo_pa_vs_ps}

In this section, we examine the interaction between the thermal reservoir and the compasslike states introduced in the previous sections. The temporal evolution of the Wigner functions associated with the PA and PS cases of these compasslike states is presented here by incorporating their components of \S{\ref{subsec:pa_case_math}}. Here, the main focus of our observations is to examine the impact of decoherence on the nonclassical phase-space structures included by the Wigner functions of these compasslike states.

First, we include the discussion related to the interaction between the PA case of our compasslike states and the thermal reservoir, which is made through the temporal evolution of the corresponding Wigner functions. This temporal evolution of the Wigner function for the PA case with terms $W^\pm(\zeta,\tau)$ and $W_{\bigotimes}(\zeta,\tau)$ provided in the expressions (\ref{eq:tempo_pa_chess}) and (\ref{eq:tempo_pa_cross}), respectively, with
\begin{align}
 W_{\bigoplus}(\zeta,\tau)=W^+(\zeta,\tau)+W^-(\zeta,\tau)
\end{align}
represents the temporal evolution of the central chessboard-like pattern, and the terms denoted by $W_{\bigotimes}(\zeta,\tau)$ represent the temporal evolution of the interference pattern appearing far from the phase-space origin, and hence, the total temporal evolution of the Wigner function of PA case yields
\begin{align}\label{eq:wig_sspa}
W_{\ket{\psi_{\text{PA}}}}(\zeta,\tau)=N^2_\text{PA}\left[W_{\bigoplus}(\zeta,\tau)+2 \operatorname{Re}\left(W_{\bigotimes}(\zeta,\tau)\right)\right],
\end{align}
which we plot in Fig.~\ref{fig:fig2}.

Similarly, the temporal evolution of the Wigner function associated to the PS case is written as
\begin{align}\label{eq:wig_ssps_tempo}
W_{\ket{\psi_{\text{PS}}}}(\zeta,\tau)=N^2_\text{PS}\left[W_{\bigoplus}(\zeta,\tau)+2 \operatorname{Re}\left(W_{\bigotimes}(\zeta,\tau)\right)\right],
\end{align}
and it is plotted in Fig.~\ref{fig:fig3}.\;The corresponding terms $W_{\bigoplus}(\zeta,\tau)=W^+(\zeta,\tau)+W^-(\zeta,\tau)$ with $W^{\pm}(\zeta,\tau)$ are provided in Eq.~(\ref{eq:tempo_ps_chess}), which shows temporal evolution of the central chessboard pattern. The second term $W_{\bigotimes}(\zeta,\tau)$ in Eq.~(\ref{eq:wig_ssps_tempo}) is supplied in Eq.~(\ref{eq:tempo_ps_cross}) depicts the temporal evolution of the interference originating far away from the phase space.

Figures~\ref{fig:fig2} and \ref{fig:fig3} present the temporal evolution of the Wigner functions in the given thermal reservoir for PA and PS cases of our compasslike states, respectively. Here, we include the situations for a zero value of average thermal photon number ($\Bar{n}=0$), also referred to as the photon-loss channel~\cite{kim1992}, and a non-zero value of average thermal photon number, $\Bar{n}\not=0$. We analyze the temporal evolution of corresponding Wigner functions in the thermal reservoir for a short- and long-range of time, i.e., for small and large $\tau$ values. Note that $\Bar{n}=0$ and $\tau\approx0$ represent the case when there is no interaction between our states and the thermal reservoir, and consequently, the Wigner functions of corresponding PA and PS cases may exhibit similar forms, as illustrated in \S\ref{subsec:non_Gaussian_states}.

First, let us consider the scenario when $\Bar{n}=0$ and $\tau = 0.01$. We present the corresponding Wigner distributions associated with PA and PS cases in Figs.~\ref{fig:fig2}(a) and \ref{fig:fig3}(a), respectively. By comparing these Wigner distributions with their equivalent non-interacting cases, which are shown in Figs.~\ref{fig:fig1}(b) and \ref{fig:fig1}(d), we observe that the interaction of the thermal reservoir with these states results in the decay of the non-classical parts of the corresponding Wigner functions. As we can see, this decay of non-classical features is readily observable for the sub-Planck structures contained in the central chessboard-like pattern. Figures~\ref{fig:fig2}(b) and \ref{fig:fig3}(b) represent the PA and PS cases, respectively, in which the average thermal photon number of the reservoir is raised to $\Bar{n}=0.5$ while the interaction time remains the same as in the previous case, i.e., $\tau=0.01$. We find that the non-classical phase-space structures are much more suppressed than in the preceding situation of zero-average thermal photon number. This shows that in the case of non-zero values of the average thermal photon number, the non-classical assets existing in the stated compasslike states degrade significantly faster than in the case of a zero-average thermal photon number.

As illustrated in Figs.~\ref{fig:fig2}(c) and \ref{fig:fig3}(c), for PA and PS cases, respectively, monitoring the interaction for a longer duration, that is, setting $\tau=0.03$ and maintaining the average thermal photon number $\Bar{n}=0.5$ as in the preceding cases, leads to the sub-Planck structures being smeared out of the phase space, and other non-classical structures are also weakened well here. Eventually, we raised the interaction duration to $\tau=0.1$ and the average thermal photon number of the reservoir to $\Bar{n}=1$, as illustrated in Figs.~\ref{fig:fig2}(d) and \ref{fig:fig3}(d), respectively, for the PA and PS cases of the corresponding compasslike states. The associated positive peak Wigner distributions for these examples show that the higher values of interaction time and the average thermal photon number of the thermal reservoir entirely remove the non-classical features from the phase space.

Moreover, for the case when the interaction time is very high, i.e., $\tau\rightarrow\infty$, the corresponding compasslike states are simply transformed into thermal states with the subsequent Wigner function, that is,
\begin{align}\label{eq:thermal_wig}
W_\text{th}(\zeta,\infty)=\frac{1}{\pi (2\Bar{n}+1)}\exp\left(-\frac{2|\zeta|^2}{(2\Bar{n}+1)}\right).
\end{align}
As we can see, the corresponding Wigner function of thermal states follows a Gaussian distribution centered at the phase-space origin and is unaffected by the squeezing parameter $r$ or the photon number $m$. For $\Bar{n}=0$, this Wigner function yields the Wigner function of a vacuum state.

To summarize, we examined the time evolution of the Wigner functions related to the compasslike states under investigation in the thermal reservoir. We observed these Wigner functions at zero and non-zero values of the average thermal photon number for both short and long timeframes. Our finding indicates that the non-classical phase-space structures found in the Wigner functions of these states are dispersed by this interaction, leading to a quicker decay of these characteristics at non-zero-average thermal photon number in the reservoir compared to the reservoir of zero-average thermal photon number. After a very long interaction, these states finally become thermal states, indicating that the given quantum states have lost all of their non-classical components due to the decoherence sparked by this interaction. For our purposes, we will focus more on the sub-Planck structures that are enclosed by the chessboard-like pattern and see how the interaction with the specified heat reservoir affects these tiny structures. The cavity parameters, such as the decay time $\tau$ and the average thermal photon number $\Bar{n}$ present in the reservoir, were previously adopted as $\tau=10^{-2}$ in experiments to achieve Fock states with significantly smaller values of $\Bar{n}$~\cite{PhysRevLett.65.976,PhysRevA.45.5193}. A more thorough explanation of how the thermal channel affects the sub-Planck structures of the present compasslike states is covered in the next section.

\begin{figure}
\centering
\includegraphics[width=0.5\textwidth]{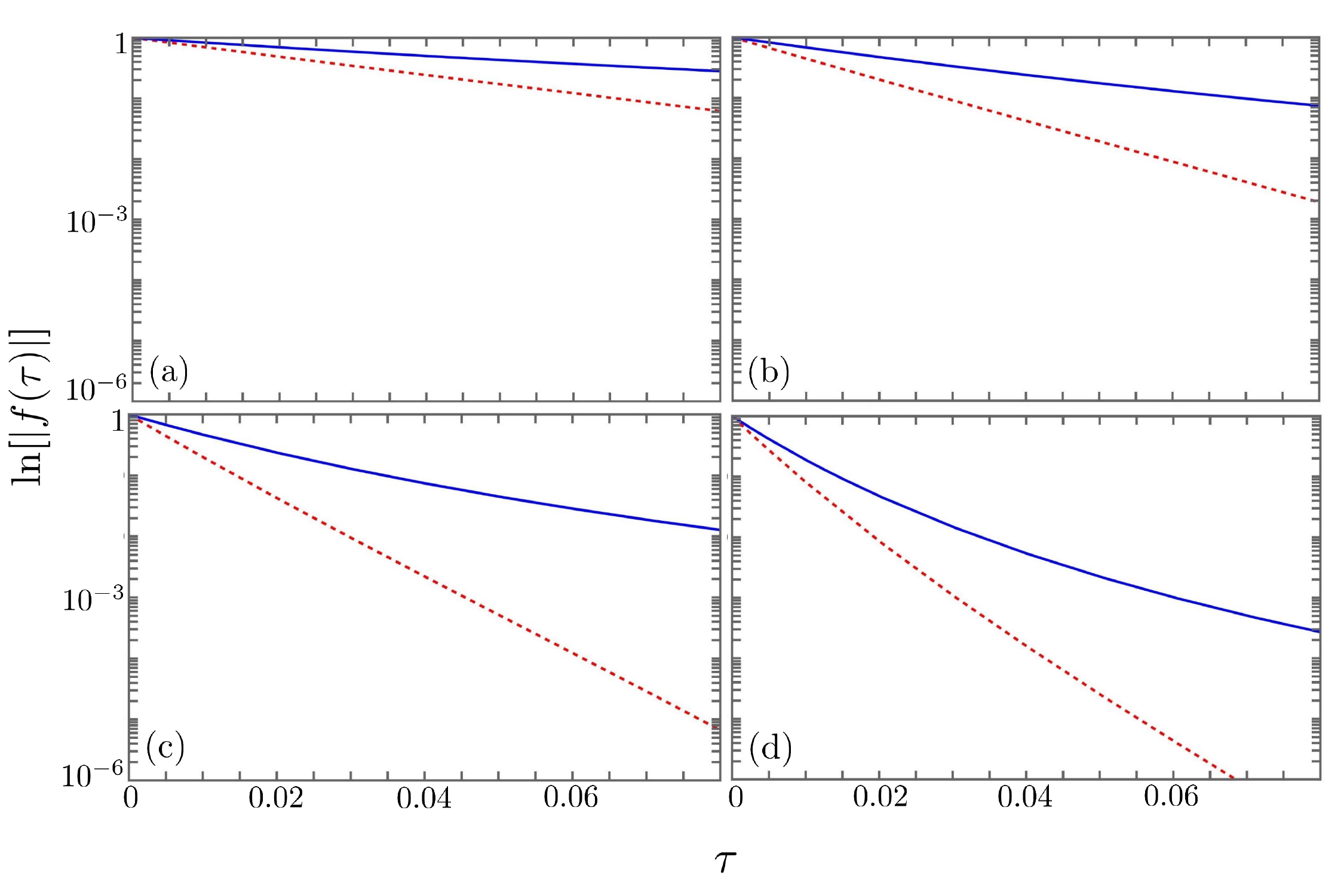}
\caption{Decay of the central sub-Planck structure for the PS case
(blue solid line) and the PA case (red dashed line): (a)~$m=5$, $r=0.5$ and $\Bar{n}=0$, (b)~$m=11$, $r=0.5$ and $\Bar{n}=0$, (c)~$m=11$, $r=0.5$ and $\Bar{n}=0.5$, and (d)~$m=11$, $r=0.8$ and $\Bar{n}=0.5$.}
\label{fig:fig4}
\end{figure}

\vspace{0.5cm}

\subsection{Deformation of core interference fringes}\label{subsec:decay_rate}
In the previous section, we discussed the temporal evolution of the Wigner distributions for the PA and PS scenarios of our compasslike states in the stated thermal reservoir. The non-classical phase-space features included by these Wigner functions degrade after the interaction with this thermal reservoir, as we can observe it by visualizing the relevant Wigner distributions. 

We now study the influence of this thermal reservoir on the sub-Planck structures contained by these PA and PS cases of the compasslike states. It has been found that sub-Planck structures of such states are the crucial and notable component~\cite{Zurek2001,naeem2023}, so it is necessary to determine how to preserve these features in order to prevent them from being destroyed after interacting with such heat channels. 
Here, we now explore the effects of the added (or subtracted) photons ($m$), the average thermal photon number ($\Bar{n}$), and the squeezing parameter ($r$) on the decay rate of these sub-Planck structures.

First, let us consider the function $f(\tau)$ that evaluates the rate at which the central sub-Planck structure of our compasslike states decays over time, that is,
\begin{equation}\label{eq:threshold_time}
f(\tau)=\frac{W_{\bigoplus}(0,\tau)}{\left|W_{\bigoplus}(0,0)\right|},
\end{equation}
where $\left|W_{\bigoplus}(0,0)\right|$ denotes the initial height of the central peak. In Fig.~\ref{fig:fig4}, we plot $\text{ln}\big(|f(\tau)|\big)$ for both the PA and PS cases of these states. We vary $m$, $r$, and $\Bar{n}$ and discuss how $f(\tau)$ behaves over $\tau$. The red dashed line represents the PA cases and the solid blue line represents the PS cases.

First, we discuss the effect of increasing amounts of added (or subtracted) photons on the decay rate of the sub-Planck structures of these states. This is demonstrated by comparing the correlative cases of PA and PS in Figs.~\ref{fig:fig4}(a) and \ref{fig:fig4}(b), where $\Bar{n}$ and $r$ are kept the same but the number $m$ representing the amount of photons being added (or subtracted) is different. It is evident that the curves in Fig.~\ref{fig:fig4}(b), which correspond to a higher number of photons added (or subtracted), i.e., $m=11$, are falling more quickly than the curves in Fig.~\ref{fig:fig4}(a), which correspond to a lower number of photons operated, i.e., $m=5$. Hence, an increase in the number of photons added or subtracted may cause the corresponding sub-Planck structure to disappear more quickly in the thermal reservoir.

Next, we investigate how varying the average thermal photon number of the thermal reservoir affects the decay rate of the sub-Planck structures of the stated compasslike states. Figures~\ref{fig:fig4}(b) and \ref{fig:fig4}(c) present the cases where the values of $m$ and $r$ are kept constant but $\Bar{n}$ is varied over each case. Note that the average thermal photon number has a direct relationship with the associated temperature of a reservoir~\cite{kim1992}; for example, $\Bar{n}=0$ corresponds to the zero-temperature thermal reservoir, and an increase in $\Bar{n}$ raises the temperature of the reservoir. Here, we show that the height of the central sub-Planck structure of both PA and PS cases falls comparatively faster for higher values of $\Bar{n}$; that is, the curve with $\Bar{n}=0.5$ presented in Fig.~\ref{fig:fig4}(c) decays faster than its previous case when $\Bar{n}=0$ as shown in Fig.~\ref{fig:fig4}(b). This indicates that an increase in the average thermal photon number of the reservoir increases the decay rate of the sub-Planck structures related to PA and PS cases, i.e., at a higher temperature of the thermal reservoir, the sub-Planck structures destroy much faster.

We now examine how the squeezing parameter associated with these compasslike states affects the decay rate of the central sub-Planck structures. This is accomplished by comparing two cases, as presented in Figs.~\ref{fig:fig4}(c) and \ref{fig:fig4}(d). Here, we vary the values of the squeezing parameter $r$, while $m$ and $\Bar{n}$ are kept constant across each case. It clearly shows that the sub-Planck structures of PA and PS cases decay faster in the specified thermal reservoir when the squeezing parameter is increased; that is, the curve shown in Fig.~\ref{fig:fig4}(d) with $r=0.8$ is decaying faster than that shown in Fig.~\ref{fig:fig4}(c) for $r=0.5$.

Let us now compare two equivalent PA and PS cases. As illustrated in Fig.~\ref{fig:fig4}, the dashed lines representing the decay rate of the central sub-Planck structure of the PA case are approaching zero faster than the decay of an equivalent PS instance. This implies that in the specified thermal reservoir, the sub-Planck structures associated with the PA cases of the defined compasslike states vanish faster than their equivalent PS cases.

\begin{table}[ht]
\centering
\caption{The numerical data for each case in Fig.~\ref{fig:fig4} and their associated temporal threshold values.}
\begin{tabular}[t]{lcccccc}
\toprule
\toprule
$i $ &\hspace{0.9cm} $m$ &\hspace{0.9cm} $r$ &\hspace{0.9cm} $\Bar{n}$ &\hspace{0.9cm}$\tau^{\text{PA}}_{\text{d}}$&\hspace{0.9cm}$\tau^{\text{PS}}_{\text{d}}$ \\
\hline
1&\hspace{0.9cm}5&\hspace{0.9cm}0.5&\hspace{0.9cm}0&\hspace{0.9cm}0.34552&\hspace{0.9cm}0.35321\\
2&\hspace{0.9cm}11&\hspace{0.9cm}0.5&\hspace{0.9cm}0&\hspace{0.9cm}0.32775&\hspace{0.9cm}0.34657\\
3&\hspace{0.9cm}11&\hspace{0.9cm}0.5&\hspace{0.9cm}0.5&\hspace{0.9cm}0.19556&\hspace{0.9cm}0.20573\\
4&\hspace{0.9cm}11&\hspace{0.9cm}0.8&\hspace{0.9cm}0.5&\hspace{0.9cm}0.17497&\hspace{0.9cm}0.20073\\
\bottomrule
\bottomrule
\end{tabular}
\label{t1}
\end{table}

In summary, we examined the decay rate of the sub-Planck structures associated with the compasslike states of the present work in the indicated thermal reservoir. We discovered that increasing the average thermal photon number of the reservoir, or the number of added or subtracted photons from these superposed states, as well as the amount of squeezing, may accelerate the decay of these sub-Planck structures. Furthermore, sub-Planck structures in the PA cases degrade faster in the thermal channel than in the PS situations.

\subsection{Approximating temporal thresholds}\label{subsec:compare_pa_vs_ps}
Here, we roughly estimate the time at which the height of the central sub-Planck structure of our compasslike states reaches its minor, i.e., $|f(\tau_\text{d})|\approx 0$, where $\tau_\text{d}$ is the information of corresponding times, and for convenience, we denote $\tau^{\text{PA}}_{\text{d}}$ and $\tau^{\text{PS}}_{\text{d}}$ as the temporal threshold values for PA and PS situations, respectively. We obtain these temporal thresholds for each of the examples given in Fig.~\ref{fig:fig4} and then provide this numerical data in Table \ref{t1} with $i=1, 2,3, 4$, corresponding to Figs.~\ref{fig:fig4} (a)-\ref{fig:fig4}(d). In the following, we compare these temporal thresholds for two comparable PA (or PS) versus PA (or PS) examples, denoted simply as PA-PA (or PS-PS). To make this comparison more qualitative, we measure relative change \cite{vartia1976relative} between two comparable situations using $\Delta=\nicefrac{(\tau^{i+1}_{\text{d}}-\tau^{i}_{\text{d}})}{|\tau^{i}_{\text{d}}|}$ with $\tau^{i}_{\text{d}}$ representing the reference value of the temporal threshold at a specific $i$ in Table \ref{t1}, while $\tau^{i+1}_{\text{d}}$ indicates the next value of the temporal threshold in the same column.
The relative change estimates the variations from one number to another and expresses the change as an increase (for $\Delta >0$) or decrease (for $\Delta<0$). In our case, using this ratio, we can roughly estimate how much faster one temporal threshold is than the other, and now we analyze each case in the following.

First, let us examine the scenario shown in Fig.~\ref{fig:fig4}(a) with $m=5$, $\Bar{n}=0$, and $r=0.5$, for which we approximate corresponding temporal thresholds and presented in Table \ref{t1} across $i=1$. For the case presented in Fig.~\ref{fig:fig4}(b) with $m=11$, $\Bar{n}=0$, and $r=0.5$, we show temporal thresholds values along $i=2$ in Table \ref{t1}. A comparison between these two cases shows that for a greater value of $m$ both $\tau^{\text{PA}}_{\text{d}}$ and $\tau^{\text{PS}}_{\text{d}}$ are noticeably reduced. In the PA-PA comparison, increasing $m=5$ to $m=11$ results in a decrease in the temporal threshold of $\Delta^\text{PA-PA}=5.14297$\%, while in the PS-PS situation, the decrease in temporal threshold is about $\Delta^\text{PS-PS}=1.8799$\%. Hence, this indicates that an increase in the number of added or subtracted photons causes the sub-Planck structures of the PA and PS instances of the presented compasslike states to decay more quickly.

Similarly, for the case represented in Fig.~\ref{fig:fig4}(c) with $m=11$, $\Bar{n}=0.5$, and $r=0.5$, we show $\tau^{\text{PA}}_{\text{d}}$ and $\tau^{\text{PS}}_{\text{d}}$ values along $i=3$ in Table \ref{t1}, and then for the case shown in Fig.~\ref{fig:fig4}(d) with $m=11$, $\Bar{n}=0.5$, and $r=0.8$, we have $\tau^{\text{PA}}_{\text{d}}$ and $\tau^{\text{PS}}_{\text{d}}$ along $i=4$. Here, the comparable situations shown along $i=2$ and $i=3$ are compared first, and we observe that the corresponding temporal thresholds are decreased. For these scenarios, we measure the relative decrease in these temporal threshold values as $\Delta^\text{PA-PA}=40.3326$\% and $\Delta^\text{PS-PS}=40.6383$\%. This implies that the sub-Planck structures of both the PA and PS cases significantly degrade when the average thermal photon number of the thermal reservoir rises. Again, we observe a decrease in the values of $\tau^{\text{PA}}_{\text{d}}$ and $\tau^{\text{PS}}_{\text{d}}$ for the cases presented along $i=3$ and $i=4$ in Table \ref{t1} with different values of $r$, and the relative decrease in the temporal threshold values are measured as $\Delta^{\text{PA-PA}}=10.5287$\% and $\Delta^{\text{PS-PS}}=2.43037$\%, indicating that the corresponding sub-Planck structures of our compasslike states decay more quickly as the squeezing parameter increases.

Finally, consider two identical PA and PS situations presented in Table \ref{t1} from $i=1$ to $i=4$ (see also Fig.~\ref{fig:fig4}). The numerical values shown in Table \ref{t1} clearly indicate that $\tau^{\text{PA}}_{\text{d}}<\tau^{\text{PS}}_{\text{d}}$ for each presented case. Furthermore, this is also clearly visible in Fig.~\ref{fig:fig4}, and hence compared to their counterparts in PS cases, we may claim that the sub-Planck structures associated with the PA cases of our compasslike states decay faster in thermal channels.

In summary, in this section, we presented qualitative research to corroborate the findings provided in \S{\ref{subsec:decay_rate}}. In the following part, we will provide a high-level description of the physical explanations for our findings.

\section{Highlights and remarks}\label{sec:conc}
In this section, we provide the key findings of the present work and their extensive physical explanations. Let us first highlight the crucial parts of our investigations. The Wigner functions of superposed photon-added and photon-subtracted SVSs may exhibit sub-Planck phase-space structures similar to the original compass states, and PA and PS cases of these compasslike states are the main constituents of this work.\;In particular, here, we investigated the interaction of these compasslike states with a thermal reservoir.\;The sub-Planck structures contained by the Wigner functions of these states (see Fig.~\ref{fig:fig1}) were the main concern of the present analysis, and we particularly observed the effect of thermal reservoir on these features (see Figs.~\ref{fig:fig2}-\ref{fig:fig4}).\;It is found that environmental decoherence aroused by the interaction of these states with the thermal reservoir results in a washout of the associated sub-Planck structures, and an increase in the quantity of the photons involved in PA and PS, or the squeezing parameter, or the average thermal photon number contained by the reservoir may lead to the faster decay of these tiny features.\;Comparatively, the sub-Planck structures of the PA case of these compasslike states are found to be more susceptible to environmental decoherence.\;However, the sub-Planck structures of these compasslike states can be preserved in the thermal reservoir for comparatively higher values of the added or subtracted photons if we set the average thermal photon number of the reservoir to nearly zero.

Now we provide a brief physical interpretation of these findings by comparing them with the previous results; we start with the well-known coherent-state superpositions that are comparable to the states of the present work. Here, we recall how changing parameters of a cat state affects its susceptibility to decoherence.\;For example, in the case of macroscopic superposition of coherent states, particularly cat states~\cite{PhysRevA.45.6570}, higher separation between two coherent states cause the enhancement of the non-classical phase-space attributes of such states~\cite{schleich1987oscillations,PhysRevA.45.6570}.\;It has been found that the deformation of a cat state in a thermal reservoir also depends on the separation between the component states; that is, higher values of the macroscopic parameter may enhance the fragility of such states against environmental decoherence~\cite{kim1992}.\;Moreover, the macroscopic parameter of a cat state is directly correlated with its average photon number and the non-classical nature~\cite{schleich1987oscillations,PhysRevA.45.6570}; that is, an increase in this parameter directly increases these two quantities.\;This means that a cat state with a greater macroscopic parameter holds a greater non-classical nature and a higher average photon number, hence being more susceptible to external decoherence~\cite{kim1992}.\;This concept directly applies to our quantum states, which we discuss in the following.

Now consider the compasslike states of the present work. These compasslike states significantly differ from cat states in the way that now the number of added or subtracted photons plays the role of the macroscopic parameter; that is, when photons are added or removed from SVSs, the nonclassical phase-space features and average photon number of these states are boosted in a manner similar to cat states~\cite{WANG2019102,Richard2014}. Interestingly, for the same number of photons added (or subtracted) to SVSs, the resulting quantum states hold different nonclassical properties and average photon number; that is, photon-added cases are always richer in these two quantities~\cite{WANG2019102,Richard2014}. This difference between these two states also has an impact on their phase-space structures; that is, the sub-Planck structures of the photon-added case are smaller than their equivalent in the photon-subtracted case, and hence the photon-added case appears to be more useful for metrological applications~\cite{naeem2023}. Holding these differences, the photon-added and photon-subtracted SVSs of the present work also behave differently in the thermal reservoir; that is, the photon-added case is highly fragile against decoherence as compared to the photon-subtracted case.

Finally, we remark that our findings may be useful for researchers interested in the generation of such states in cavities, and perhaps our observations may also help in the development of a technique to protect such states in noisy channels.\;Protecting quantum states from the environment has always been an important topic~\cite{Jeannic2018,Pan2023}.\;How to preserve the specified quantum states of this study in a noisy environment is an open question that can be addressed in future research.
\section*{Acknowledgement}
This work is supported by the Natural Science Foundation of Jiangsu Province (Grant No. BK20231320) and the National Natural Science Foundation of China (Grant No. 12174157). NA acknowledges Chuchi for her helpful feedback on the work. MA acknowledges support from the Khalifa University of Science and Technology under Award No. FSU-2023-014.
\vspace{0.1cm}

\onecolumngrid

\appendix
\section{Multitude of Mathematics}\label{appendix_b}

In the following, we present a number of the substitutions employed in the \S\ref{subsec:pa_case_math}.

\begin{align}
&\Bar{\zeta}^\pm=\zeta \cosh (r)\mp\zeta^* \sinh (r),A^{\pm}=\frac{4\text{e}^{-2\kappa t}}{\midpoint{T}^2}\left(\text{e}^{-2\tau}+2\midpoint{T}\left(\cosh (2r)\right)^{\pm1}+\midpoint{T}^2\text{e}^{2\tau}\right),B^{\pm}_1=\frac{8\text{e}^{-4\tau}}{\midpoint{T}^3}\Big(|\zeta|^2+\midpoint{T}\text{e}^{2\tau}|\Bar{\zeta}^\mp|^2\Big),\\&C^\pm_1=\frac{8\text{i}\text{e}^{-3\tau}\sqrt{\pm 2\coth r}}{\midpoint{T}^2}\Big(\Bar{\zeta}^{\pm *}+\midpoint{T}\text{e}^{2\tau}\Bar{\zeta}^{\mp *}\Big),
D^\pm_1=-\frac{8\text{i}\text{e}^{-3\tau}\sqrt{\pm 2\coth (r)}}{\midpoint{T}^2}\Big(\Bar{\zeta}^{\pm}+\midpoint{T}\text{e}^{2\tau}\Bar{\zeta}^{\mp}\Big),\\&E_1=\frac{16 \text{e}^{-2\tau}\cosh^2 (r)}{\midpoint{T}}, G^\pm_1=\pm 16 \coth r\left(1+\frac{\text{e}^{-2\tau}\cosh (2 r)}{\midpoint{T}}\right),\chi_1=E_1-A^+,\Theta_1=\frac{1}{2\sqrt{A^+\chi_1}},\\&\Lambda^\pm=\pm 2\sech (2r)|\zeta|^2+\big(\zeta^2-\zeta^{*2}\big)\tanh(2r),B^\pm_2=\frac{\text{e}^{-4\tau}}{\midpoint{T}^3}\left(8|\zeta|^2\pm 4\midpoint{T}\text{e}^{2\tau}\Lambda^\pm\right),\\&C_2=-\frac{8\text{i}\text{e}^{-3\tau}\Omega}{\midpoint{T}^2}\left(\Bar{\zeta}^{+ *}+\midpoint{T}\text{e}^{2\tau}\Bar{\zeta}^{- *}\right),D_2=\frac{8\text{e}^{-3\tau}\Omega}{\midpoint{T}^2}\left(\Bar{\zeta}^{-}+\midpoint{T}\text{e}^{2\tau}\Bar{\zeta}^{+}\right),E_2=\frac{4\text{e}^{-2\tau}\Omega^2 \sinh(2r)}{\midpoint{T}},\\&G_2=-\frac{8\text{i}\Omega^2}{\midpoint{T}}\big(\text{e}^{-2\tau}+\midpoint{T}\cosh(2r)\big),\chi_2=E_2-A^-,\Theta_2=\frac{1}{2\sqrt{A^-\chi_2}},\\&C^\pm_3=\frac{8\text{i}\text{e}^{-3\tau}\sqrt{\pm 2\tanh (r)}}{\midpoint{T}^2}\Big(\Bar{\zeta}^{\pm *}+\midpoint{T}\text{e}^{2\tau}\Bar{\zeta}^{\mp *}\Big),D^\pm_3=-\frac{8\text{i}\text{e}^{-3\tau}\sqrt{\pm 2\tanh (r)}}{\midpoint{T}^2}\Big(\Bar{\zeta}^{\pm}+\midpoint{T}\text{e}^{2\tau}\Bar{\zeta}^{\mp}\Big),\\&E_3=\frac{16 \text{e}^{-2\tau}\sinh^2 (r)}{\midpoint{T}},G^\pm_3=\pm 16 \tanh (r)\left(1+\frac{\text{e}^{-2\tau}\cosh (2r)}{\midpoint{T}}\right),\chi_3=E_3-A^+,\Theta_3=\frac{1}{2\sqrt{A^+\chi_3}},\\&C_4=-\frac{8\text{i}\text{e}^{-3\tau}\omega}{\midpoint{T}^2}\left(\Bar{\zeta}^{- *}+\midpoint{T}\text{e}^{2\tau}\Bar{\zeta}^{+ *}\right),D_4=\frac{8\text{e}^{-3\tau}\omega}{\midpoint{T}^2}\left(\Bar{\zeta}^{+}+\midpoint{T}\text{e}^{2\tau}\Bar{\zeta}^{-}\right),E_4=\frac{4\text{e}^{-2\tau}\omega^2 \sinh (2r)}{\midpoint{T}},\\&G_4=-\frac{8\text{i}\omega^2}{\midpoint{T}}\big(\text{e}^{-2\tau}+\midpoint{T}\cosh(2r)\big),\chi_4=E_4-A^-,\Theta_4=\frac{1}{2\sqrt{A^-\chi_4}}.
\end{align}

As additional information, we now give a few of important identities for solving mathematical problems involving the temporal evolution of the Wigner functions in the present case. To remove the $\mathrm{e}^{\gamma s t}$ terms from our complex exponential, we apply the following sum series:

\begin{align}\label{eq:st_rid}
\exp(\gamma_1 s+\gamma_2 t+\gamma_3 st)
=\sum_{l=0}^{\infty}\frac{\gamma_3^{l}}{l!}\frac{\partial^{2l}}{\partial \gamma_1^{l}\partial \gamma_2^{l}}\left[\exp\left(\gamma_1 s+ \gamma_2 t\right)\right].
\end{align}
Notice the generating function of the Hermite polynomial, and its recursive relation, we have
\begin{equation}
    H_{n}(x)=\big. \frac{\partial^{n}}{\partial s^{n}}\exp\left(2xs-s^2\right)\big|_{s=0},\,\frac{\mathrm{d}^{l}}{\mathrm{d}x^{l}}H_{n}(x)=\frac{2^{l}n!}{(n-l)!}H_{n-l}(x).
    \label{eq:eq_hermite}
\end{equation}

The following integral formula is mainly used
\begin{align}\label{eq:int_1}
        &\int^{\infty}_{-\infty}\mathrm{d}^2 \beta \exp{\bigg[A |\beta|^2+B \beta+C \beta^*+D \beta^2+E\beta^{* 2}}\bigg]=\frac{\pi}{\sqrt{A^2-4 D E}}\exp{\bigg[\frac{-A B C+B^2 E+C^2 D}{A^2-4 D E}}\bigg].
    \end{align}
\twocolumngrid
\bibliography{References}
\end{document}